This is a "cut-down" version of the paper, ontaining all of the text but only some of the figures. The full set of figures is available via anonymous ftp from `suna.amtp.liv.ac.uk` as `/pub/pss/fullpaper.ps.Z` . Login as '`ftp`' and send your email address as password. The figures are available as compressed PostScript files. The full LaTeX source is also there. Don't forget to set the transfer mode to binary!

Paddy Spencer.

# Topography of the cooled $\mathcal{O}(3)$ vacuum

' "and what is the use of a book," thought Alice, "without pictures...?" ' [1].

**P.S. Spencer and C. Michael**
DAMTP, University of Liverpool, Liverpool L69 3BX, UK

**Abstract**

We investigate the *topography* of the $\mathcal{O}(3)$ vacuum after various degrees of under-relaxed cooling, using the naïve action density and geometric and field theoretical definitions of the topological charge density. The results are presented graphically.

## 1 Introduction

The $\mathcal{O}(3)$ model has been widely studied [2] in 2 dimensions because of its similarities to non-Abelian gauge theories in 4 dimensions: it is asymptotically free, becomes non-perturbative in the infra-red regime and has a non-trivial topology due to the homotopy class $\pi_2(S_2) = \mathbb{Z}$ of windings from $S_2 \to S_2$ . The $\mathcal{O}(3)$ model is well understood in some respects—for instance an exact expression for the mass gap is known [3]. It is thus a natural candidate for a theory which one can try and understand in terms of the vacuum structure. In this paper we explore the nature of the vacuum via numerical simulation in Euclidean time. In order to focus on important features in the vacuum, we use a procedure to smooth out local fluctuations. A cooling algorithm has been used previously to achieve this. The basic idea is that a local smoothing of the fields should preserve long range features such as instantons. We explore this by varying the details of the cooling procedure and checking whether the resulting topological charge distribution is universal.

The main result we present in this paper is a series of figures showing the action and topological charge densities after cooling. In another paper [4], we analyse these distributions in terms of models—for instance the distributions of sizes of topological objects (instantons and anti-instantons) were calculated and compared with a distribution relation derived from a modification of Coleman's argument for the $SU(2)$ case [5].

The paper is organised as follows: in section 2 we outline the formalism of the $\mathcal{O}(3)$ model, give two definitions of the lattice topological charge, and discuss their suitability. We also give formulae relevant to single-instanton solutions. In section 3 we detail the cooling algorithm we used and discuss the different effects that different degrees of cooling have upon an initial, uncooled configuration. We then discuss a possible criterion for quantifying the degree of cooling required to remove the short-range fluctuations. Finally, in section 5 we show examples of the vacuum obtained from different initial configurations, using the method given in section 3.

## 2 The Model.

The continuum $2d$-$\mathcal{O}(3)$ Euclidean action $S_E$ is defined as:

$$S_E = \frac{1}{2g^2} \int d^2x (\partial_\mu \phi)^2 \tag{1}$$

with $\phi$ a 3-component vector and the constraint $\phi^2 = 1$. On the lattice the simplest discretization gives:

$$\begin{aligned} S_L &= \frac{1}{2g^2} \sum_x \sum_\mu (\phi_{x+\mu} - \phi_x)^2 \\ &= \frac{1}{g^2} \sum_x (2 - \phi_F(x) \cdot \phi_x) \end{aligned} \tag{2}$$

with $\phi_F$ defined by:

$$\phi_F(x) = \sum_\mu \phi_{x+\mu}. \tag{3}$$

The continuum topological charge $Q^T$ is given by:

$$Q^T_f = \frac{1}{8\pi} \int \varepsilon_{\mu\nu} \varepsilon_{ijk} \phi_i \partial_\mu \phi_j \partial_\nu \phi_k d^2x \tag{4}$$

and is an integer. However, if this 'field theoretical' definition is naïvely discretised, it returns non-integer values and acquires a renormalisation factor [6].

There is also a 'geometric' lattice definition of $Q^T$ based on the mapping $S_2 \rightarrow S_2$ of the $\mathcal{O}(3)$ fields to the lattice triangles formed by the fields around a plaquette. The contribution to the total charge from a site $x^*$ on the dual lattice is given by [7–9]:

$$Q^T_g(x^*) = \frac{1}{4\pi}((\sigma A)(\phi_1, \phi_2, \phi_3) + (\sigma A)(\phi_3, \phi_4, \phi_1)) \tag{5}$$

where

$$\sigma(\phi_1, \phi_2, \phi_3) = \mathrm{sign}(\phi_1 \cdot \phi_2 \times \phi_3) \ , \tag{6}$$

$A(\phi_1, \phi_2, \phi_3)$ is the area of the spherical triangle on the unit sphere mapped out by the $\phi$ fields, and the subscripts $1, 2, 3, 4$ refer to the corners of a plaquette labelled anti-clockwise, as in fig. 1. Obviously, there is a choice of triangulation, and for configurations with small action the two equivalent possibilities give equal contributions to the topological charge but for other configurations the difference in the contribution can be $\pm 1$.

There are arguments against the use of such geometric definitions, particularly in calculations of the topological susceptibility, $\chi^T$, in $CP^{N-1}$ theories with small values of $N$. These arguments arise because these lattice models are plagued by dislocations: unphysical configurations that contribute to $\chi^T$. Campostrini et al have shown [10] that for values of $N = 10$ and larger, the geometric formulation gives a sensible definition of $\chi^T$, but fails to do so for lower values of $N$.

When $\phi$, which is purely real, is written in terms of the complex projective fields $\omega$, $\overline{\omega}$:

$$\phi_1 = \frac{\omega + \overline{\omega}}{\omega\overline{\omega} + 1}\ ,\ \phi_2 = \frac{\omega - \overline{\omega}}{i(\omega\overline{\omega} + 1)}\ ,\ \phi_1 = \frac{\omega\overline{\omega} - 1}{\omega\overline{\omega} + 1} \tag{7}$$
$$\omega = \frac{\phi_1 + i\phi_2}{1 - \phi_3}$$

the continuum action and topological charge densities are seen to be closely related to one another [11]. Specifically, in the case of (anti-)instanton solutions, a field configuration representing a single instanton of size $\rho$ at position $r$ can be constructed via

$$\omega = \frac{\rho}{z - r}\ ,\ \text{with}\ \ z = x + it. \tag{8}$$

An instanton so constructed is circular in the continuum, but, being an extended object, is only circular on the lattice if $\rho$ is small compared to $L$, the dimension of the lattice.

With $\phi$ and $\omega$ as in eqs. 7 and 8 above, the action and topological charge densities become:

$$S(z)\ ,\ Q(z) \propto \frac{\rho^2}{(\rho^2 + \mid z - r \mid^2)^2}. \tag{9}$$

A similar relation holds for multi-instanton configurations. It should be noted that, for a general field configuration, the action is positive definite, but the topological charge is not, having contributions from both anti- and instantons for which the charge has opposite signs. A single (anti-)instanton solution has continuum action

$$S_I = \frac{4\pi}{g^2} \tag{10}$$

In general a multi instanton–anti-instanton solution has action

$$S \geq \frac{4\pi}{g^2} \mid Q^T \mid \tag{11}$$

# 3 Cooling.

Our aim is to study the topography—the distribution of topological charge in a configuration. As can be seen from fig. 2, little if anything can be deduced about a configuration's topography from an uncooled configuration—a thermalised, i.e. simulated at non-zero $g^2$, configuration contains too many short-range fluctuations, and the longer scale structure is obscured.

Cooling is a local update procedure, and as such is used with the intention of removing the short scale fluctuations while leaving intact the larger scale structure of interest. We employed an under-relaxed cooling:

$$\phi_x \rightarrow \alpha\phi_x + \phi_F(x) \tag{12}$$

(normalised to $\phi_x^2 = 1$ and with $\phi_F(x)$ as in eq. 3 above). The cooling usually performed (as in e.g. [6]) is a Metropolis-style accept-reject cooling at $g^2 = 0$, equivalent to only accepting new field values that *lower* the action. Cooling using $\alpha = 0$ in eq. 12 is equivalent to a Heatbath algorithm, or a Metropolis update where each site is updated very many times before moving on to the next site. Taking non-zero $\alpha$ has, for our purposes, a number of advantages: firstly, our implementation is deterministic which is computationally less demanding; secondly, the parameter $\alpha$, together with the number of cooling sweeps used, allows the nature of the cooling performed to be controlled. Provided the cooling algorithm is implemented sensibly, a higher $\alpha$-value requires a larger number of cooling sweeps to reach the same $N$-sector as can be reached by fewer sweeps at a lower $\alpha$-value. Ideally a checkerboard approach should be adopted, as this updates only non-interacting sites on each sweep and is thus maximally insensitive to the order in which field variables are cooled.

We work on a $64^2$ lattice at $g^2 = 0.8$ using the naïve discretisation of the action given above in eq. 2. Configurations are separated by 1000 Monte Carlo sweeps, each consisting of 1 heatbath and 9 over-relaxation updates. The correlation length at this value of $g^2$, we measure to be $\xi = 3.82(3)$. Studies at other correlation lengths will be presented elsewhere [4]. Some examples of the cooled vacuum are shown in section 5.

It is convenient to define

$$N_{f,g} = \sum_x \mid Q_{f,g}(x) \mid \tag{13}$$

Generally, different degrees of cooling will result in different stabilities versus number of cooling sweeps for the different $N$-sectors, as can be seen from fig. 3. The more gentle cooling, that with a larger $\alpha$, will tend to isolate the various $N$-sectors for longer. Plots such as fig. 3 have appeared before in the literature and the plateaux were described by multi-instanton configurations; the jumps between the plateaux were ascribed to the annihilation of an (anti-)instanton by the cooling process. This can be observed by following the history of the configuration in computer time as it is cooled. What happens is the topological object is shrunk by the cooling process from its uncooled size, and eventually, under prolonged cooling, it is reduced to a point and then destroyed, at which stage in the cooling a jump occurs in $N$.

We now investigate whether variations in the cooling algorithm give different topological distributions. It is necessary to look beyond plots such as fig. 3 to study this. Indeed, each $N$-sector is not constructed uniquely from instanton–anti-instanton solutions—a configuration with two instantons and one anti-instanton will have the same value of $N$ as one with one instanton and two anti-instantons, but surely lies in a different topological sector. Similarly, two different cooling processes need not lead to the same topological sector. We have observed configurations which, when cooled under differing degrees of cooling, lead to final configurations where the topological objects removed by the cooling process are not the same for the different coolings, and consequently, although $N$ is identical for these resultant field configurations, they may lie in different topological sectors.

There are further consequences to this. Prolonged cooling of a configuration should be avoided, as the structure remaining after a large amount of cooling is more an artefact of the cooling process than any large-scale physics that was present before the cooling. As a criterion for how much cooling should be performed to remove the short-range fluctuations without adversely affecting the physics present in a configuration, we investigated the amount of cooling required to annihilate single-instanton configurations generated via eq. 8. Some results are shown in fig. 4. As can be seen there is a linear relation between the value of $\alpha$ and the number of sweeps needed to remove an instanton of a given size. This is equivalent to cooling until a particular value of $\overline{S}/(VS_I)$ is attained, where $V$ is the volume of the lattice and the bar denotes average taken over configurations. It appears to be the case that, whilst matching the number of cooling sweeps to the value of $\alpha$ in this way leads to *broadly* the same final configuration, the configurations become (visually) much more alike if more sweeps of a gentler cooling is used. For example, figs. 5a to 5e are the final configurations for the same initial configuration where the amount and degree of cooling have been chosen from relationships similar to fig. 4. In this case the criterion was that an instanton of size $\rho = 2$ on a $64^2$ lattice be annihilated by the cooling—equivalent to $\overline{S}/(VS_I) = 0.0017(3)$ calculated on 145 configurations. For this value of $g^2$, the various degrees of cooling employed appear to produce very nearly identical end configurations, for non-zero $\alpha$. As our aim is to disturb the large-scale structure much less than the short-range fluctuations, we therefore chose to cool subsequent configurations at $\alpha = 2$ for 90 sweeps, as the differences between figs. 5a and 5b are much more significant than those between fig. 5b and the rest; clearly fewer sweeps are desirable in order to minimise the extent to which the long-range objects are affected.

It has been suggested in [12] that, particularly in gauge theories, lattice instantons are unstable under cooling because the lattice instanton action is smaller than in the continuum. They propose a so-called "over-improved" action that stabilises the instantons. Certainly our $\mathcal{O}(3)$ lattice instantons have $S_I^L \leq S_I^{\mathrm{cont}}$, and are shrunk by the cooling process. We concentrate on relatively large topological objects under relatively mild cooling so we expect little bias. It would be interesting to explore the use of over-improved actions to improve the stability of instantons under cooling. Another approach, the "perfect" action, has been proposed [13] and this would allow instanton effects to be studied closer to the continuum limit.

## 4 Conclusions

Fig. 6 shows the absolute topological charge, calculated using both the geometric and field theoretic definitions and the normalised action, as one configuration is cooled at $\alpha = 2$, and fig. 7 shows $S/S_I$ and $Q_{f,g}$ for the same configuration. It is interesting to note, in the light of the discussion in section 2, that after a certain amount of cooling, $N_g$ and $S/S_I$ track each other much more closely than $N_f$. This relation between $S$ and $Q^T$ will be useful in calculations in $SU(N)$ gauge theories, where the definition of $Q_L^T$ is non-trivial and any calculation of the topological charge is much more computationally intensive than a similar calculation of $S_L$.

As a demonstration of points made above, we present figs. 8a–8l. These show action and geometric topological charge densities for different vacuum configurations, each generated by cooling using $\alpha = 2$, for 90 sweeps. This combination was chosen to match the criterion defined in section 3 above. The criterion here was again to annihilate a $\rho = 2$ instanton on a $64 \times 64$ lattice—equivalent to $\overline{S}/(VS_I) = 0.0017(3)$ as stated earlier. The close relationship between $S(x,t)$ and $Q_g(x,t)$ is readily apparent, and is shown more clearly in fig. 9 which shows contour plots of the last vacuum example.

In this paper we show that cooling configurations can give reliable information about the $\mathcal{O}(3)$ vacuum topography. Moreover, after such cooling the topological and action densities track each other closely so that there is sufficient information contained within the action density for topological calculations that do not depend on the sign of the topological objects. In particular, a study of the topological susceptibility is plagued by the difficulty of determining the contribution of small-size topological objects on a lattice. By looking at the full distribution itself, it is possible to focus on those properties which should be well determined by traditional lattice methods—namely the distribution of instantons above a cetain size. We return to this in [4].

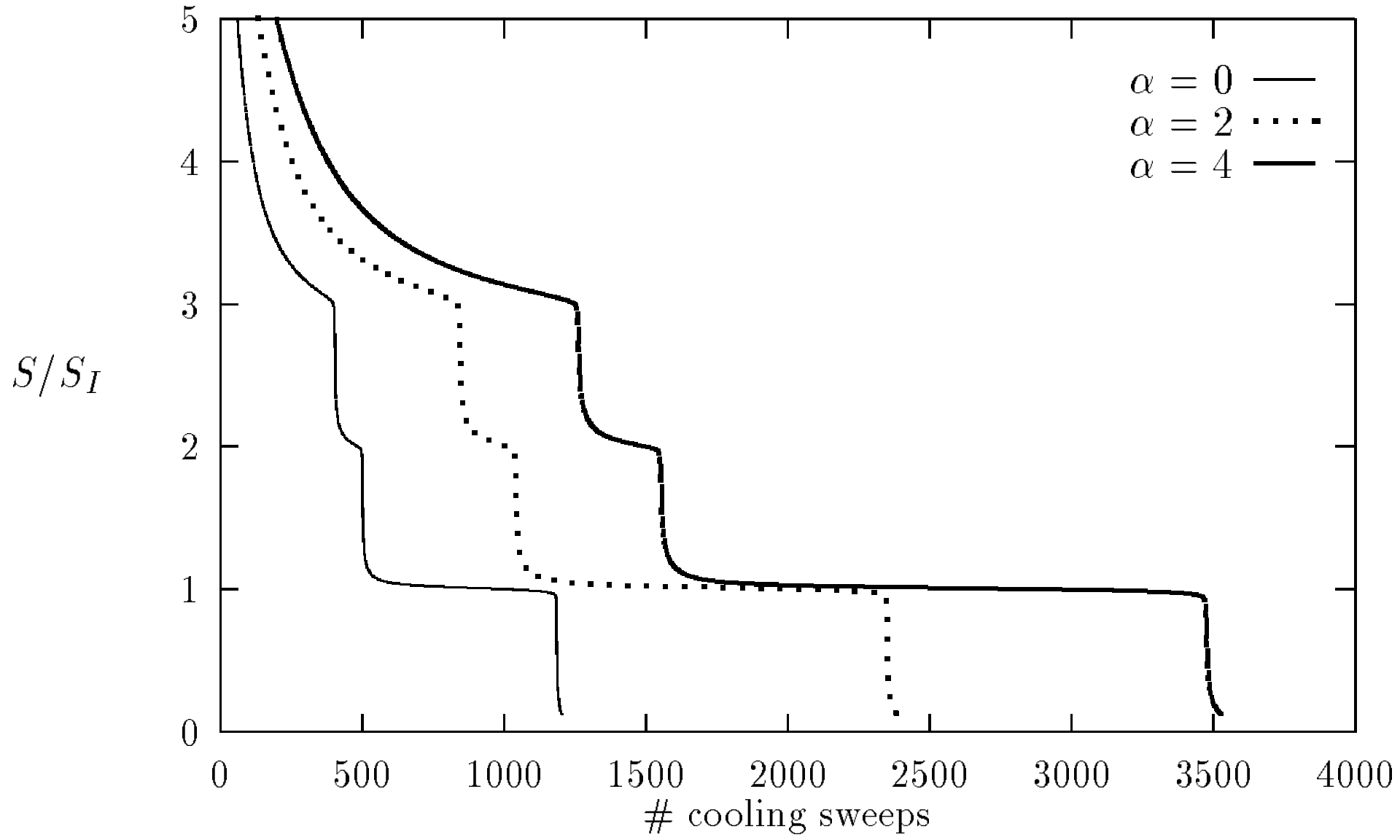

Figure 3: $S/S_I$, for a configuration as it is cooled with $\alpha = 0, 2, 4$.

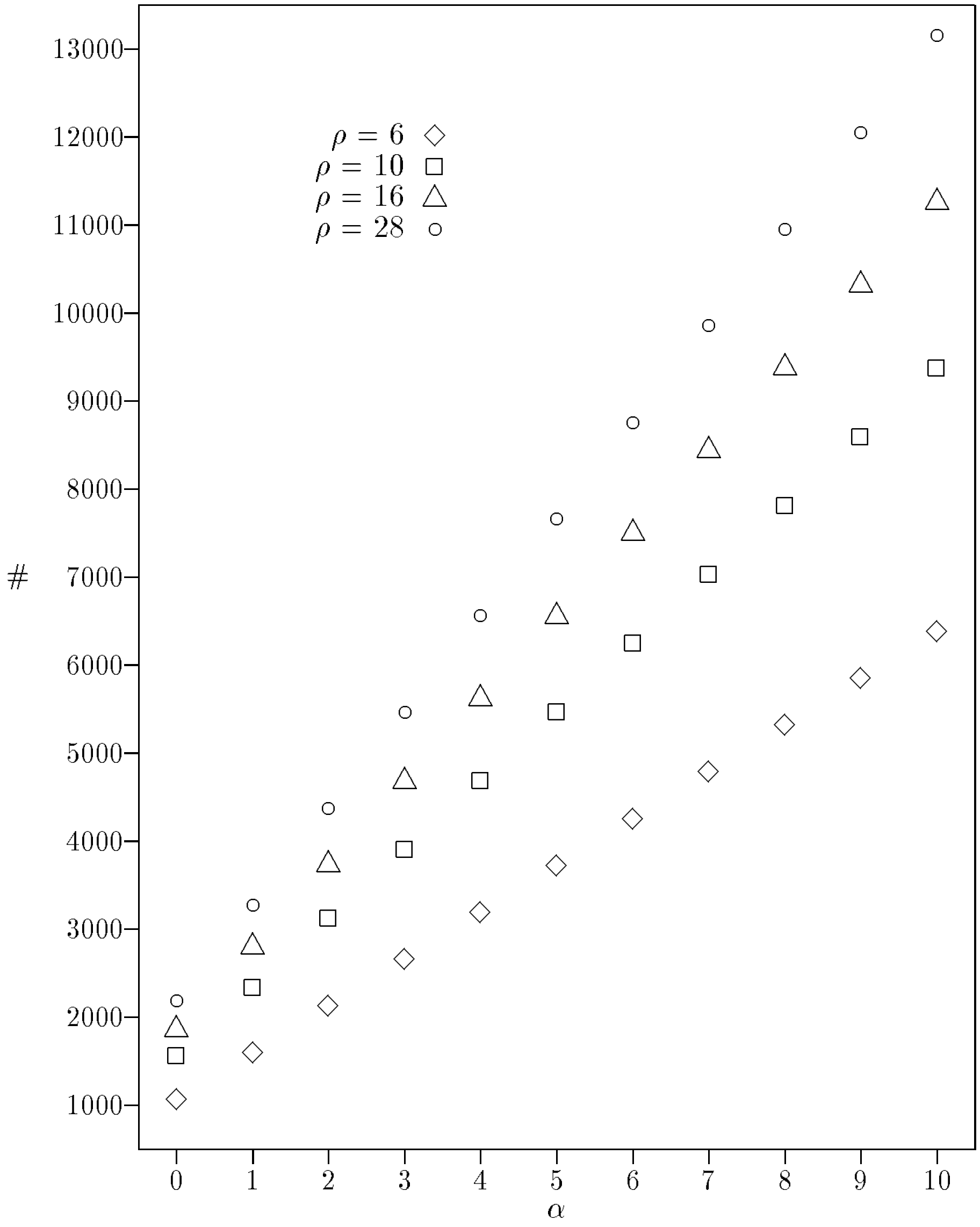

Figure 4: The number of cooling sweeps required to annihilate a single, size $\rho$ (as defined in eq. 8) instanton configuration at a given value of $\alpha$. Data calculated on a $64 \times 64$ lattice with the instanton positioned at the centre of the lattice.

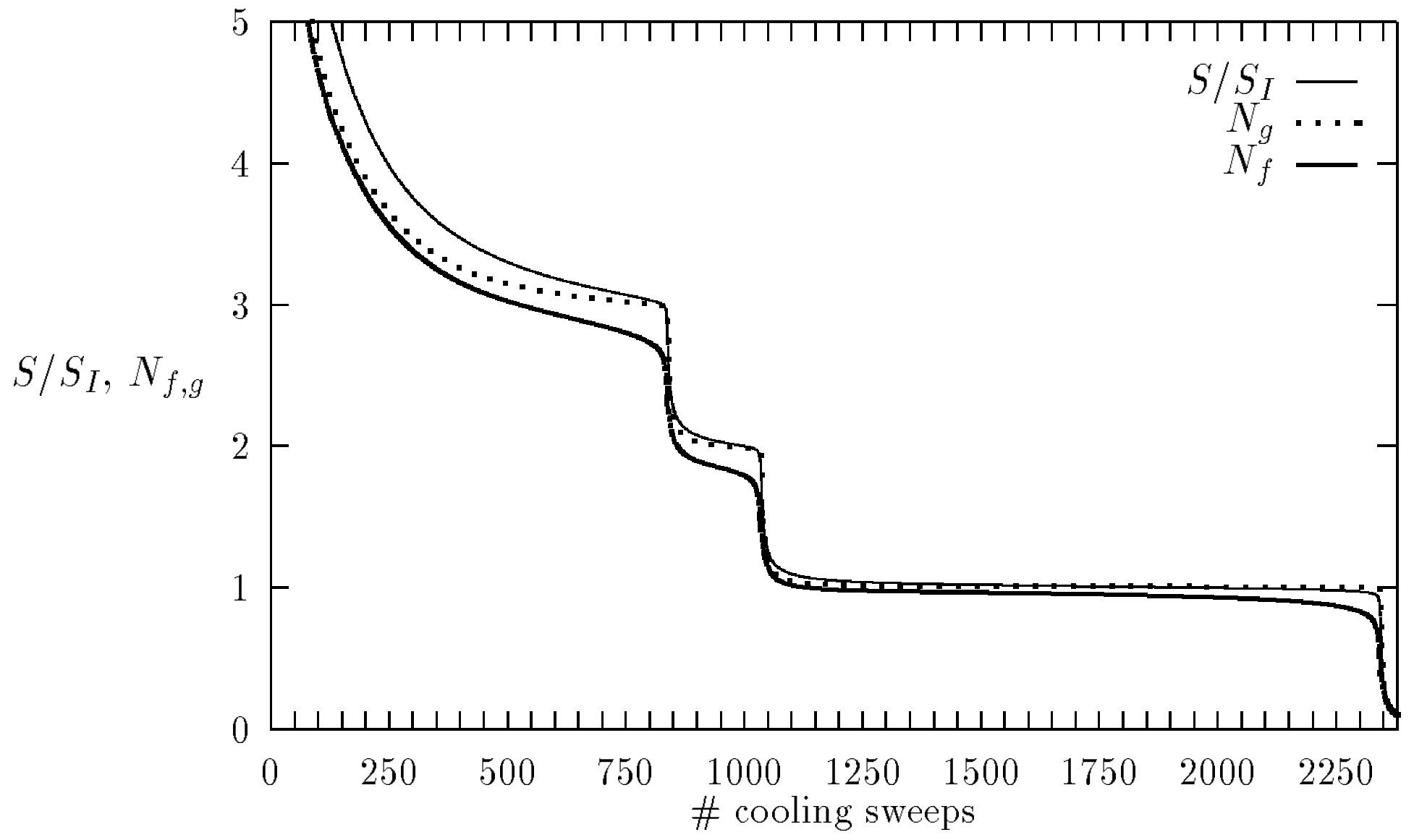

Figure 6: $N_{f,g}$ as defined in eq. 13, and the normalised action $S/S_I$. These results were obtained by cooling with $\alpha = 2$.

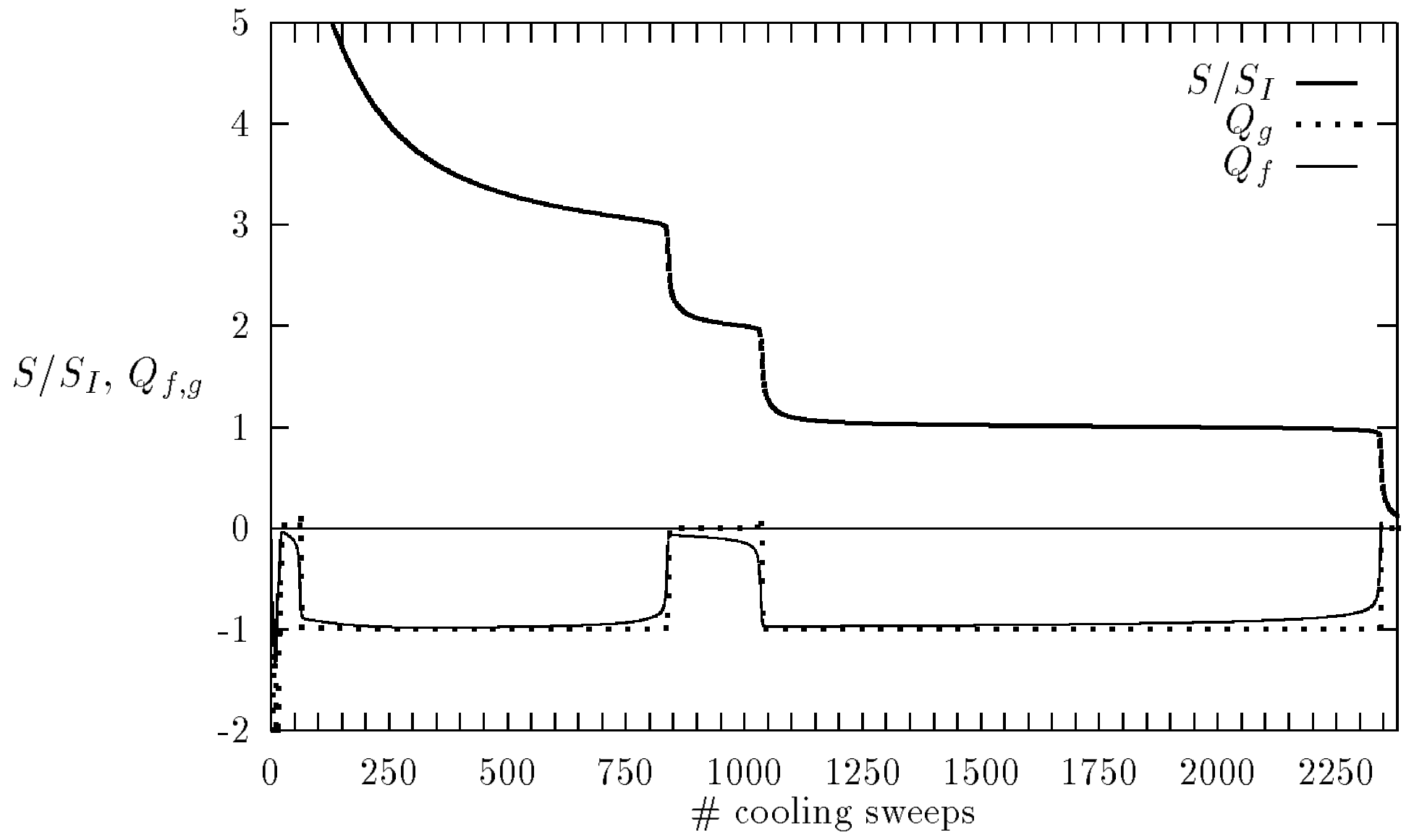

Figure 7: $Q_{f,g}$ and the normalised action $S/S_I$. These results were obtained by the same process that generated fig. 6.

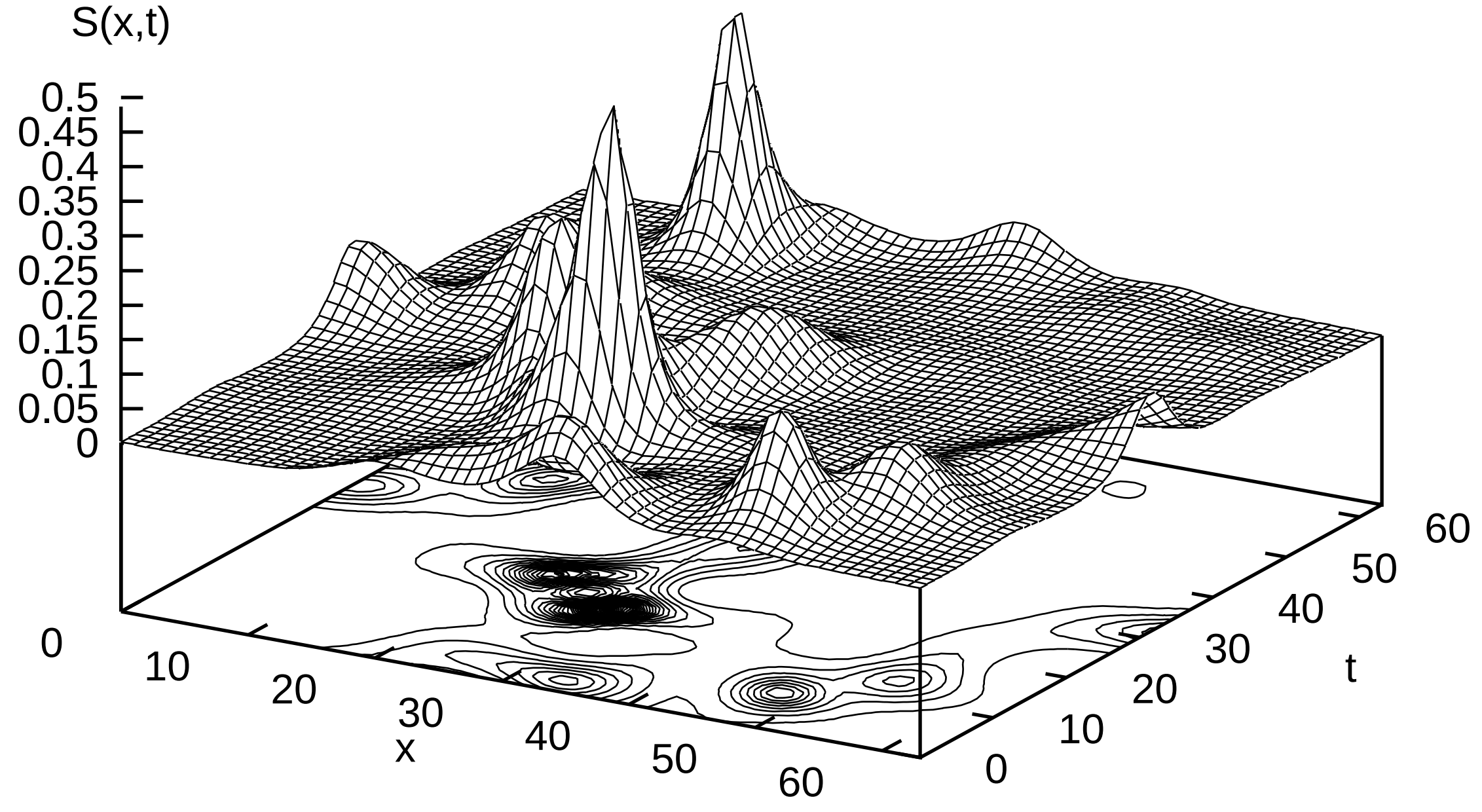

Figure 8a: This and the following figures show $S(x,t)$ and $Q_g(x,t)$ for different configurations, each obtained from an uncooled configuration such as in fig. 2 by 90 sweeps at $\alpha = 2$

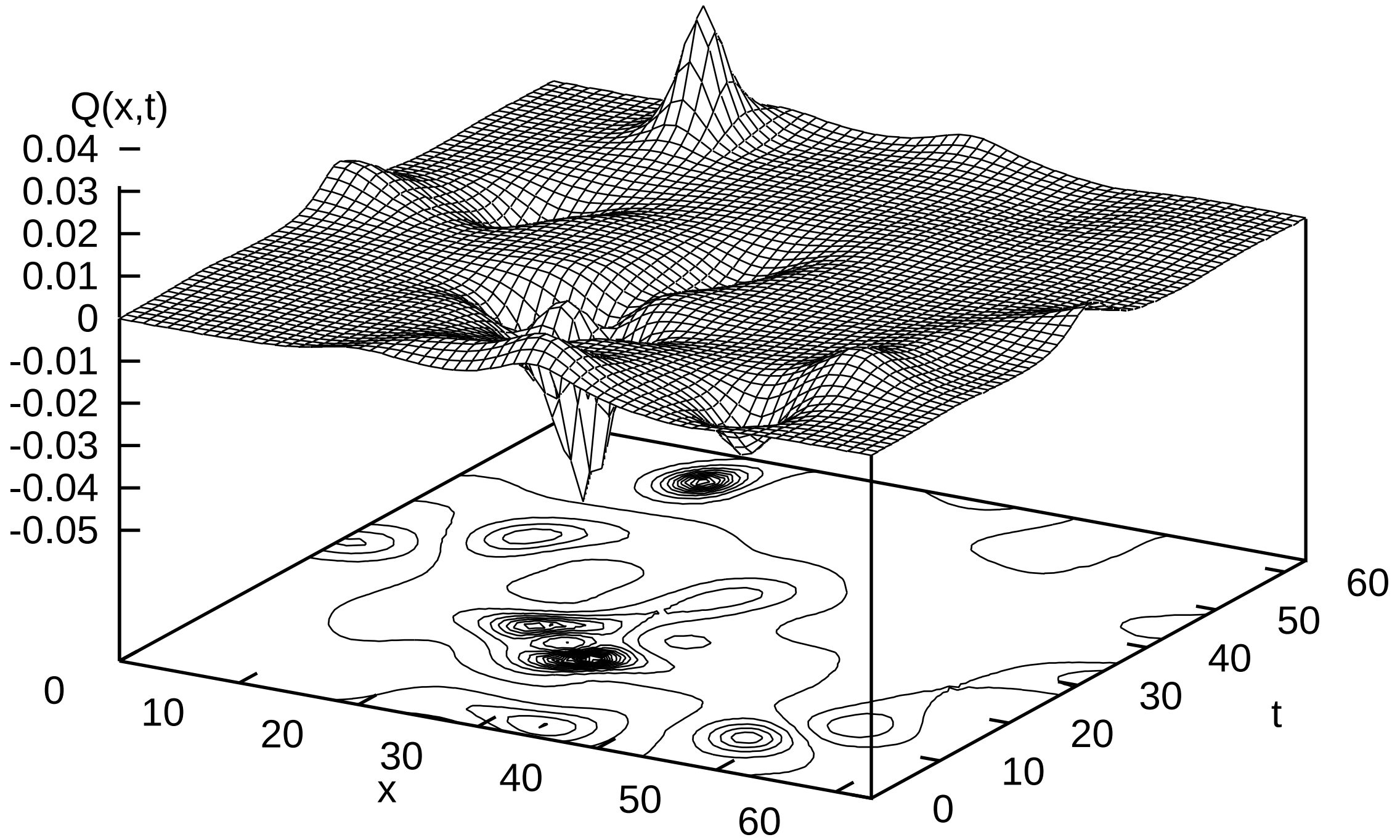

Figure 8b: $Q_g(x,t)$ for fig. 8a.